\documentclass[12pt,a4paper]{article}

\usepackage{amssymb}
\usepackage[dvips]{color}
\usepackage[centerlast,footnotesize]{caption2}
\usepackage{float}

\usepackage{multirow}
\usepackage{booktabs}
\usepackage{latexsym}
\usepackage{amsmath}
\usepackage{amsbsy}
\usepackage{epsfig}
\usepackage{graphicx}
\usepackage{latexsym}
\usepackage{euscript}
\usepackage{amsfonts}
\usepackage{mathtools}
\usepackage{braket}
\usepackage{simplewick}
\usepackage{amsmath}
\usepackage{amssymb}
\usepackage{dsfont}
\usepackage{slashed}
\usepackage{verbatim}

\newcommand{\gapprox}{\raisebox{-0.5ex}{$\
\stackrel{\textstyle>}{\textstyle\sim}\ $}}

\usepackage{graphicx}
\newcommand{\One}{1\kern-4.5pt1}

\newcommand{\be}{\begin{equation}}
\newcommand{\ee}{\end{equation}}

\def\lesim{${\lower 2pt\hbox{$\scriptstyle
<$}\atop\raise 4pt\hbox{$\scriptstyle\sim$}}$} 
\def\grsim{${\lower2pt\hbox{$\scriptstyle >$} \atop\raise4pt\hbox 
{$\scriptstyle\sim$}}$} 

\begin{document}
\begin{center}
\begin{flushright}
April 2025
\end{flushright}
\vskip 10mm
{\LARGE
Critical Behaviour in the Single Flavor\\ Thirring Model in 2+1$d$ with
\vspace{0.15cm}\\Wilson
Kernel Domain Wall Fermions
}
\vskip 0.3 cm
{\bf Simon Hands$^a$  and Jude Worthy$^b$}
\vskip 0.3 cm
{\em $^b$ Department of Mathematical Sciences, University of Liverpool,\\
Liverpool L69 3BX, United Kingdom.}
\vskip 0.3 cm
{\em $^b$ Department of Physics, College of Science, Swansea University,\\
Singleton Park, Swansea SA2 8PP, United Kingdom.}
\end{center}

\vskip 1.3 cm
\noindent
{\bf Abstract:} 
We present results of a lattice field theory simulation of the 2+1$d$ Thirring
model with $N=1$ fermion flavors, using domain wall fermions. The model exhibits
a U(2) symmetry-breaking phase transition with the  potential to define a UV-stable
renormalisation group fixed point. The novelty is the
replacement of the Shamir kernel used in all previous work with the Wilson
kernel, improving the action particularly with respect to the 
$L_s\to\infty$ limit needed to recover U(2), now under much
better control.
Auxiliary field ensembles generated on $16^3\times24$ with varying self-interaction strength $g^2$
and bare mass $m$ are
used to measure the bilinear condensate order parameter
$\langle\bar\psi i\gamma_3\psi\rangle$ with domain wall separations as large as
$L_s=120$. The resulting $L_s\to\infty$ extrapolation is used to fit an empirical
equation of state modelling spontaneous symmetry breaking as $m\to0$. The
fit is remarkably stable and compelling, with  the
fitted critical exponents $\beta_m\simeq2.4$, $\delta\simeq1.3$ differing markedly
from previous estimates. The associated susceptibility exhibits a mass hierarchy
in line with physical expectations, again unlike previous estimates.
Schwinger-Dyson equation (SDE) solutions of the Thirring model exploiting a hidden
local symmetry in the action are reviewed, and analytic predictions
presented for
the exponents. In contrast to  all previous lattice studies, the
universal characteristics of the critical point 
revealed qualitatively resemble the SDE predictions.

\vspace{0.5cm}


\newpage
\section{Introduction}
The $2+1d$ Thirring model describes relativistic fermions interacting via a
local contact term between conserved currents, with
continuum Lagrangian density
\begin{equation}
{\cal L}_{\rm
cont}=\sum_{i=1}^N\bar\psi_i(\partial\!\!\!/\,+m)\psi_i+{g^2\over2N}\sum_i(\bar\psi_i\gamma_\mu\psi_i)^2.
\label{eq:Lcont}
\end{equation}
Here $N$ is the number of flavors each described by a 4-component spinor. Since there are
two Dirac matrices $\gamma_3,\gamma_5$ which anti-commute with the kinetic term, 
for $m=0$
(\ref{eq:Lcont}) has a symmetry under U(2$N$) global flavor rotations which is
broken to U($N)\otimes$U($N$) by a bilinear mass term such as $\bar\psi\psi$,
$i\bar\psi\gamma_{3,5}\psi$. For sufficiently strong interaction strength this
breaking can occur spontaneously, and it is believed that at fixed $N$ the coupling $g_c^2$ 
marks a UV-stable renormalisation group fixed point where an interacting
continuum limit may be found. 

As further elaborated in Sec.~\ref{sec:SDE} below, since 
small values of $N$ are expected to be the most important and accordingly {\it
a priori\/} no small parameter available, non-perturbative methods are
mandatory when studying the symmetry-breaking dynamics. Analytic approaches
applied include Schwinger-Dyson equations
(SDE)~\cite{Gomes:1990ed,Itoh:1994cr,Ebihara:1994wm,Sugiura:1996xk} and Functional
Renormalisation Group (FRG)~\cite{Gies:2010st,Janssen:2012pq,Gehring:2015vja}. 
There have also been several approaches based on
numerical simulation of lattice field theory -- see \cite{Hands:2021eyc} for a
recent review. Interest in the problem has been renewed with the realisation
that for a strongly-coupled problem the lattice formulation should ideally faithfully respect the
correct U(2$N$) symmetry at a microscopic level, leading to recent studies using both Domain Wall
(DWF) and SLAC fermions~\cite{Wipf:2022hqd}.

The DWF approach has been explored in a series of
papers~\cite{Hands:2016foa,Hands:2018vrd,Hands:2020itv,Hands:2022fhq}. In order
to recover U($2N$) it is necessary to take the limit $L_s\to\infty$ where $L_s$
is the separation of the domain walls in a fictitious ``third'' direction $x_3$.
While the nature of the limit is quite well
understood~\cite{Hands:2015qha,Hands:2015dyp}, achieving it with
sufficient numerical control has proved challenging, in part because the
lattice formulation to be reviewed in Sec.~\ref{sec:form} below employs
non-unitary link fields $1+iA_\mu$ making the fermion matrix $D_{\rm DWF}[A]$ rather ill-conditioned.
Recent attempts with $N=1$ and $L_s\leq48$ found evidence that while U(2) is
recovered only rather slowly as $L_s\to\infty$ it is indeed
spontaneously broken at sufficiently strong coupling, implying that the critical
$N_c$ above which the symmetry is unbroken at all couplings is bounded by $N_c>1$.
The associated equation of state (EoS, specified in (\ref{eq:EoS}) below) 
in the vicinity of the transition is characterised by two
critical exponents $\beta_m\approx0.3$,
$\delta\approx4$~\cite{Hands:2018vrd,Hands:2020itv}.

In this paper we apply an improved DWF operator to the problem, replacing the
previously-used Shamir kernel $\gamma_3D_W(2+D_W)^{-1}$
of the associated $2+1d$ overlap operator $D_{\rm ov}$ with the
better-behaved Wilson kernel $\gamma_3D_W$. Operationally, this is implemented by altering the
definition of the derivative  in the third direction in the DWF operator, as
described in Sec.~\ref{sec:form} below. Details of the algorithm
are set out in \cite{Worthy:2024lmc}. The Wilson kernel's eigenvalue spectrum is
bounded from above, and pilot studies~\cite{Hands:2023rsd,Worthy:2024lmc} show it to be much
better conditioned than the Shamir kernel in the symmetry broken phase, 
at the cost of slightly lower RHMC
acceptance. It also gives superior convergence to
U(2$N$)-symmetry as $L_s\to\infty$, as measured by the 
residual in the Ginsparg-Wilson relation~\cite{Hands:2020itv}.

In this paper we restrict attention to the model with $N=1$.
By studying the order parameter $\langle\bar\psi\psi(g^2,m)\rangle$ on spacetime
volume $16^3$ extrapolated
to $L_s\to\infty$ we have found the  Wilson kernel formulation yields a more
compelling fit to an empirical critical EoS than obtained
previously, with significant impact on the critical parameters.
The critical coupling $g_c^2$ is shifted to a much weaker value than the
estimate of \cite{Hands:2020itv}, and the
exponents modified to $\beta_m\gapprox2$, $\delta\approx1.3$.
The emerging picture of criticality is much more similar to the predictions of
SDE~\cite{Sugiura:1996xk} than that suggested by any previous lattice study.

The rest of the paper is organised as follows. Sec.~\ref{sec:form} reviews
the definition of the lattice Thirring model with DWF, which employs a bosonic
auxiliary field $A_\mu$ defined on the links of the spacetime lattice, and
outlines the simulation methods. Numerical results are presented in
Sec.~\ref{sec:results}, including: details of the $L_s$ extrapolation of
condensate measurements made on ensembles $\{A_\mu\}$ generated at fixed $L_s$;
the impact of varying $L_s$ in the ensemble generation; fits to the resulting
EoS to extract critical parameters; and the associated order parameter
susceptibility $\chi_\ell=\partial\langle\bar\psi\psi\rangle/\partial m$. Pointers to
previous results obtained using Shamir kernel are given for comparison: the Wilson
kernel yields a more robust extrapolation and the EoS fit is
remarkably stable. Moreover in contrast to previous findings $\chi_\ell$
increases as $m\to0$, matching expectations based on the EoS. In
Sec.~\ref{sec:SDE} we change tack, reviewing a description of Thirring
criticality based on solving SDEs, using an approach which exploits a hidden
local symmetry in (\ref{eq:Lcont}) to predict $N_c\simeq4.32$. However, the
limits $N\to N_c$ and $g^2\to g_c^2$ are found not to commute; 
since $\langle\bar\psi\psi\rangle$ should be a
function of state this suggests this analysis is not complete.
Much of this material already exists in the
literature~\cite{Itoh:1994cr, Sugiura:1996xk}, but here special attention is
paid to deriving critical exponents. Finally Sec.~\ref{sec:disc}
discusses the numerical results in this context, and speculates on the
$N$-dependence of the universal scaling at the fixed points with $N<N_c$.

\section{Formulation \& Methodology}
\label{sec:form}
Fermion fields $\Psi(x,s),\bar\Psi(x,s)$ with four spinor components are defined on the sites of an
$L^3\times L_s$ lattice, where $L^3$ determines the spacetime volume and $L_s$
the domain wall separation, via the bilinear form
\begin{eqnarray}
\bar\Psi D_{DWF}\Psi&\equiv&
\sum_{x,y}\sum_{s,s^\prime}
\bar\Psi(x,s)\Bigl[\delta_{s,s^\prime}D_{Wx,y}[A]
+D_{3x,y;s,s^\prime}[A]\Bigr]\Psi(y,s^\prime)\\
&+&im\sum_{x,y}\left(-\bar\Psi(x,L_s)(1-D_W)_{x,y}P_
-\Psi(y,1)+\bar\Psi(x,1)(1-D_W)_{x,y}P_+\Psi(y,L_s)\right),\nonumber
\label{eq:bilinear}
\end{eqnarray}
where $m$ is the bare mass, $P_\pm={1\over2}(1\pm\gamma_3)$,
\begin{equation}
D_{Wx,y}[A]\equiv-{1\over2}\sum_{\mu=0,1,2}\Bigl[(1-\gamma_\mu)(1+iA_\mu(x))
\delta_{x+\hat\mu,y}+(1+\gamma_\mu)(1-iA_\mu(x))\delta_{x-\hat\mu,y}\Bigr]+(3-M)\delta_{x,y},
\end{equation}
with $M$ the domain wall height, and
\begin{equation}
D_{3x,y;s,s^\prime}[A]=\Bigl[(D_{W,x,y}[A]-1)P_-\delta_{s+1,s^\prime}(1-\delta_{s^\prime,L_s})
+(D_{Wx,y}[A]-1)P_+\delta_{s-1,s^\prime}(1-\delta_{s^\prime,1})\Bigr]+\delta_{s,s^\prime}.
\label{eq:D3}
\end{equation}
Here $A_{\mu}(x)$ is a real three-dimensional auxiliary boson field defined on the link
between $x$ and $x+\hat\mu$, whose fluctuations are moderated by the gaussian
form $S_{\rm bose}=\sum_{x,\mu}A^2_\mu(x)/2g^2$. Integration over $A$ yields
four-fermi interactions expressible as a contact interaction between
conserved but non-local fermion currents~\cite{Hands:2022fhq} depending on $\Psi,\bar\Psi$ throughout
the bulk. The model is completed by
specifying fields $\psi,\bar\psi$ living in the Thirring model's 2+1$d$ target space
in terms of $\Psi,\bar\Psi$ on the domain walls:
\begin{equation}
\psi(x)=P_-\Psi(x,1)+P_+\Psi(x,L_s);\;\;\;
\bar\psi(x)=\bar\Psi(y,L_s)(1-D_W)_{y,x}P_-+\bar\Psi(y,1)(1-D_W)_{y,x}P_+,
\end{equation}
whence the mass term in (\ref{eq:bilinear}) is seen to be equivalent to
$im\bar\psi\gamma_3\psi$.

A key property of the Wilson operator is its $\gamma_3$-hermiticity
$\gamma_3D_W\gamma_3=D_W^\dagger$. In the large-$L_s$ limit it is possible to
show~\cite{Hands:2015dyp} that $D_{DWF}$ is equivalent to the 2+1$d$ overlap
operator\footnote{For $L_s$ finite Wilson-kernel $D_{DWF}$ is equivalent to 
a truncated overlap operator.}
\begin{equation}
D_{\rm ov}={1\over2}\left[(1+im\gamma_3)+{D_W\over\sqrt{D_W^\dagger
D_W}}(1-im\gamma_3)\right],
\label{eq:overlap}
\end{equation}
in which case the U(2) symmetry of the continuum Thirring model is realised in
the Ginsparg-Wilson sense. The appearance of ${\rm sgn}(\gamma_3D_W)$ in
(\ref{eq:overlap}) underlies the designation ``Wilson kernel''.
The equivalence holds in the sense that $\lim_{L_s\to\infty}{\rm
det}[D_{DWFh}^{-1}(m=1)D_{DWF}(m)]={\rm det}D_{\rm ov}(m)$, where for $D_{DWHh}$
the antihermitian mass term in (\ref{eq:bilinear}) is replaced by the more familiar
hermitian $m\bar\psi\psi$~\cite{Hands:2015dyp}.

Monte Carlo simulations of the model were performed on a
$16^3\times L_s$ lattice using the RHMC algorithm set out in
\cite{Hands:2018vrd,Hands:2020itv}, with domain wall height $Ma=1$, but with the
Shamir kernel employed in those studies replaced by the Wilson kernel
(\ref{eq:D3}) as described in~\cite{Worthy:2024lmc}. Initially 
auxiliary field ensembles were generated with $L_s=24$, which were then used to
estimate the bilinear order parameter for U(2) symmetry breaking
$i\langle\bar\psi\gamma_3\psi\rangle$ using $L_s=24,32,48,72,96$ and 120, chosen to
facilitate the $L_s\to\infty$ extrapolation. The bilinear was estimated on each
configuration using 10 independent gaussian noise vectors for each of the components
$-\bar\Psi(L_s)(1-D_W)P_-\Psi(1),\bar\Psi(1)(1-D_W)P_+\Psi(L_s)$. The required matrix
inversions were performed using a conjugate gradient routine with stopping
residual $10^{-5}$ per vector element. The non-compact nature of the auxiliary
$A_\mu$ makes the inversion numerically challenging; the number of CG iterations
required for measurement on $L_s=24$(120) rises from $\sim4500$ ($\sim5500$) at the weakest
coupling $g^2$ and largest fermion mass $m$ explored to $\sim12000$ ($\sim14000$) at the
strongest coupling and smallest mass. Over the same parameter range the RHMC acceptance rate on $L_s=24$
varied from $\sim$82\% to $\sim$65\%.

\begin{figure}[htbp]
\begin{center}
\includegraphics[scale=0.80]{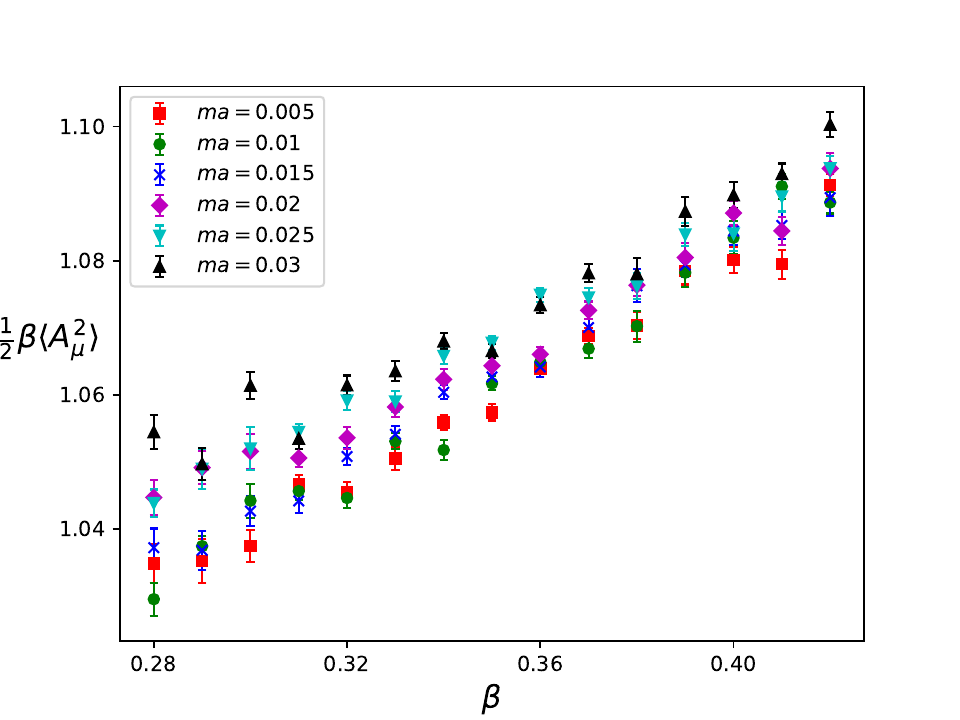}
\end{center}
\caption{Auxiliary boson action density vs. coupling $\beta$ from RHMC
production runs on $16^3\times24$, averaged using a binsize of 20 to mitigate
autocorrelations.}
\label{fig:Sbose}
\end{figure}
Pilot studies using the Wilson kernel action on a $12^3$ spacetime
lattice~\cite{Hands:2023rsd,Worthy:2024lmc} suggest a U(2) symmetry breaking
transition in the region $\beta\sim0.30$ -- 0.35, where the dimensionless
coupling $\beta\equiv ag^{-2}$. Accordingly we generated a total of 90
auxiliary ensembles with $\beta\in[0.28,0.29,\ldots,0.42]$ and bare mass
$ma\in[0.005,0.01,\ldots,0.03]$. Configurations were stored following every 5 RHMC
trajectories, with random trajectory length drawn from a Poisson distribution with mean length 1.0; in all
cases at least 100 configurations were generated for each parameter set,
with
300 configurations generated for the region $\beta\in[0.31,0.37]$ where, based on
visual inspection of the $L_s=24$ data, the order parameter fluctuations 
were largest.

Fig.~\ref{fig:Sbose} shows the boson action density estimated over the entire
parameter set. With admittedly large statistical fluctuations, this
quantity appears to vary smoothly and evenly within the confines of the
parameter range explored.
Similarly to what was found for the bulk formulation with Shamir
kernel, interaction with the fermions reduces $S_{\rm
bose}$ below its non-interacting value $S_{\rm bose}^{\rm free}={3\over2}$,
the effect increasing as $m\to0$. Compared to Shamir kernel the departure from free-field
behaviour is significantly enhanced, as can be seen by comparison with Fig.~13 of
\cite{Hands:2020itv}.  There is
no sign of the non-monotonic behaviour seen in that study,
which hinted at complicated behaviour in the
strong-coupling limit. It is also striking that within the critical region identified
in studies using the Shamir kernel $S_{\rm bose}$ is a {\em decreasing} function
of $\beta$, in contrast to the behaviour shown in Fig.~\ref{fig:Sbose}. 
The physical interpretation of $S_{\rm bose}$ in the bulk
formulation is not
transparent~\cite{Hands:2018vrd}, and it is clear far greater statistics would
be needed to obtain it with precision. The effect of generating the auxiliary
ensemble with $L_s>24$ will be addressed below.

\section{Results \& Analysis}
\label{sec:results}
\subsection{Extrapolation $L_s\to\infty$}
\label{sec:extrap}
Our strategy is a partially-quenched one:\footnote{The optimised DWF studied in
\cite{Hands:2023rsd,Worthy:2024lmc} which use an improved approximation to the
signum function in (\ref{eq:overlap}) 
were found to be prohibitively expensive on the spacetime volume used in
this study.} we approach the limit
$L_s\to\infty$ by performing measurements in the valence sector for
$L_s=24,\ldots,120$. First, in order to assess U(2) symmetry recovery in
this limit we define  a residual $\delta_h$: 
\begin{equation}
\delta_h(L_s)={1\over2}{\rm Im}\left\langle
\bar\Psi(1)(1-D_W)P_+\Psi(L_s)+\bar\Psi(L_s)(1-D_W)P_-\Psi(1)\right\rangle.
\end{equation}
\begin{figure}[htbp]
\begin{center}
\includegraphics[scale=0.48]{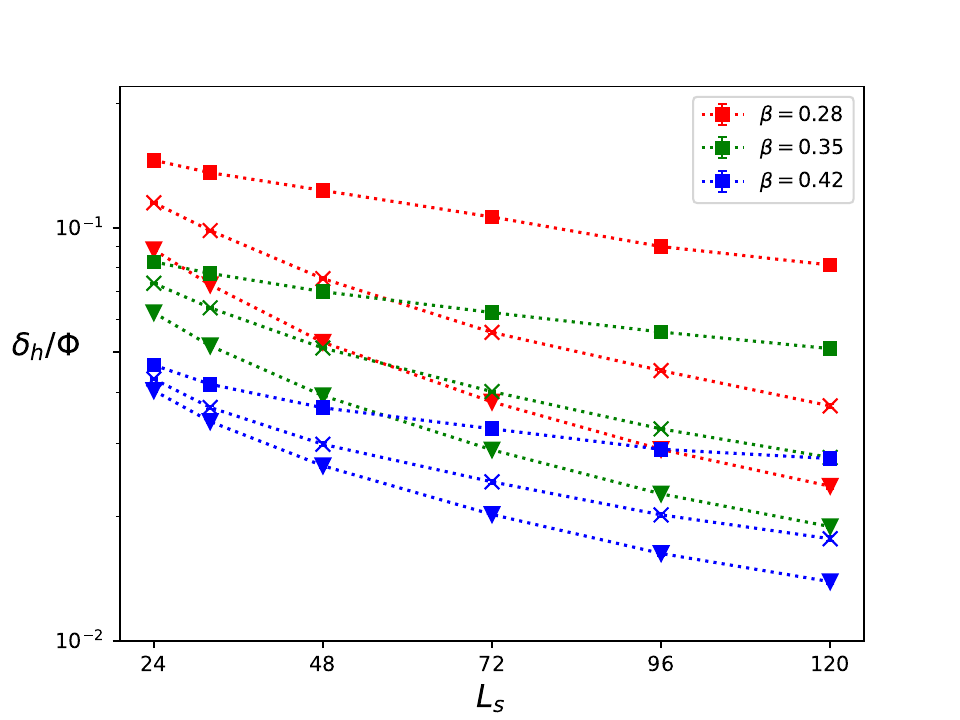}
\includegraphics[scale=0.48]{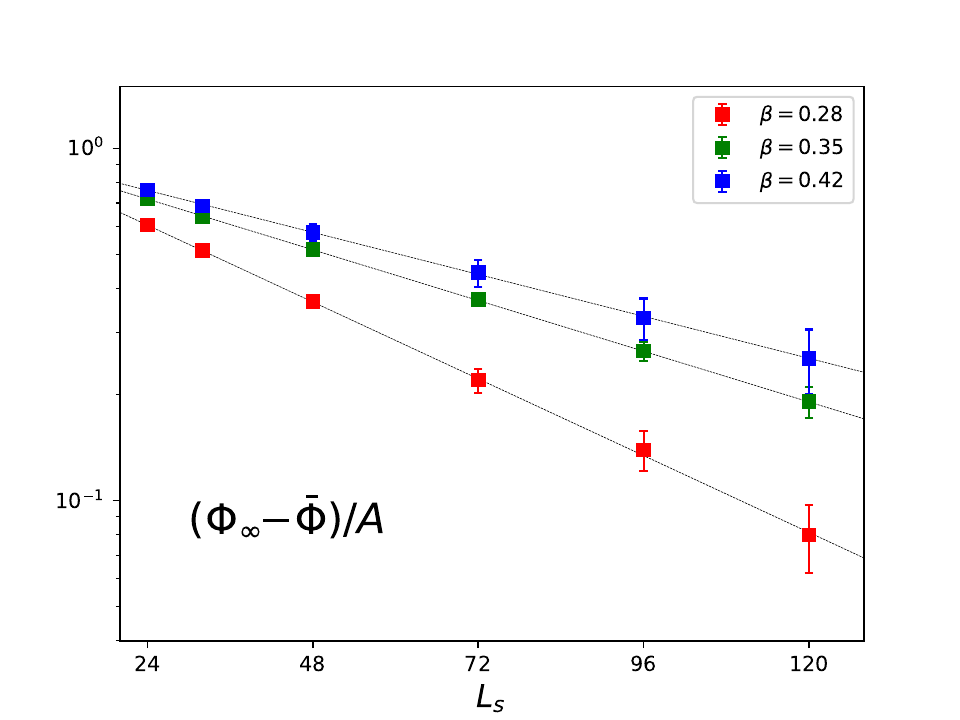}
\end{center}
\caption{(Left) Ratio of the residual $\delta_h(L_s)$ to the condensate $\Phi(L_s)$ plotted on a log scale 
for three representative couplings and $ma=0.005$ ($\Box$), 0.015 ($\times$),
0.025 ($\nabla$); (Right) $(\Phi_\infty-\bar\Phi)/A$ vs. $L_s$ on the same 
scale for the same three couplings at $ma=0.005$, together with a fit to
(\ref{eq:extrapolate}).}
\label{fig:delta}
\end{figure}
In \cite{Hands:2015qha} it was demonstrated that $\delta_h$ is an effective
proxy\footnote{At least in a model where the bose fields vary
smoothly and compactly, such as QED$_3$.} for the splitting between the
condensates $\langle\bar\psi\psi\rangle$ and
$i\langle\bar\psi\gamma_3\psi\rangle,i\langle\bar\psi\gamma_5\psi\rangle$, which
must vanish in a U(2)-symmetric  theory. The left panel of 
Fig.~\ref{fig:delta} shows $\delta_h/\Phi$ (where $\Phi$ stands as a shorthand
for the order parameter $i\langle\bar\psi\gamma_3\psi\rangle$) as a function of
$L_s$ for three representative couplings. The data fall in three groups,
ranked in decreasing magnitude as the  mass $ma$ increases. The ratio falls monotonically
consistent with U(2) restoration in the $L_s\to\infty$
limit, but Fig.~\ref{fig:delta} suggests the decay is slower than exponential. In fact, replotting with log
scales on both axes suggests the decay is also faster than a power law. We
deduce that for this parameter set and range of accessible $L_s$ the residual 
$\delta_h$ has a mild coupling dependence and
is governed by more than one, indeed possibly several, scale(s). Corresponding
data obtained with the Shamir kernel is shown in Fig.~12 of
\cite{Hands:2018vrd}.

Using measurements taken on $N_{L_s}$ different $L_s$ with independent noisy
sources,
we extrapolate results to the $L_s\to\infty$ limit using an exponential {\em
Ansatz}~\cite{Hands:2018vrd}:
\begin{equation}
\Phi(L_s)=\Phi_\infty-A\exp(-\Delta L_s),
\label{eq:extrapolate}
\end{equation}
where fit parameters $A,\Delta$ and $\Phi_\infty$ all depend on $m$ and $\beta$.
Since $\Phi(L_s)$ data are taken on the same underlying auxiliary field
ensembles, the fit takes correlations into account by minimising 
$\chi^2=(\Phi^{\rm fit}-\bar\Phi)C^{-1}(\Phi^{\rm fit}-\bar\Phi)$ where 
$\Phi^{\rm fit}(L_s;\Phi_\infty,A,\Delta)$ is given by (\ref{eq:extrapolate}),
$\bar\Phi$ denotes the average over the dataset at fixed $L_s$ and the
$N_{L_s}\times N_{L_s}$ covariance matrix is
\begin{equation}
C_{L_{s1},L_{s2}}={1\over{N_{\rm dat}(N_{\rm dat}-1)}}\sum_{i=1}^{N_{\rm dat}}
(\Phi_{i,L_{s1}}-\bar\Phi_{L_{s1}})(\Phi_{i,L_{s2}}-\bar\Phi_{L_{s2}}).
\end{equation}
\begin{figure}[htbp]
\begin{center}
\includegraphics[scale=0.48]{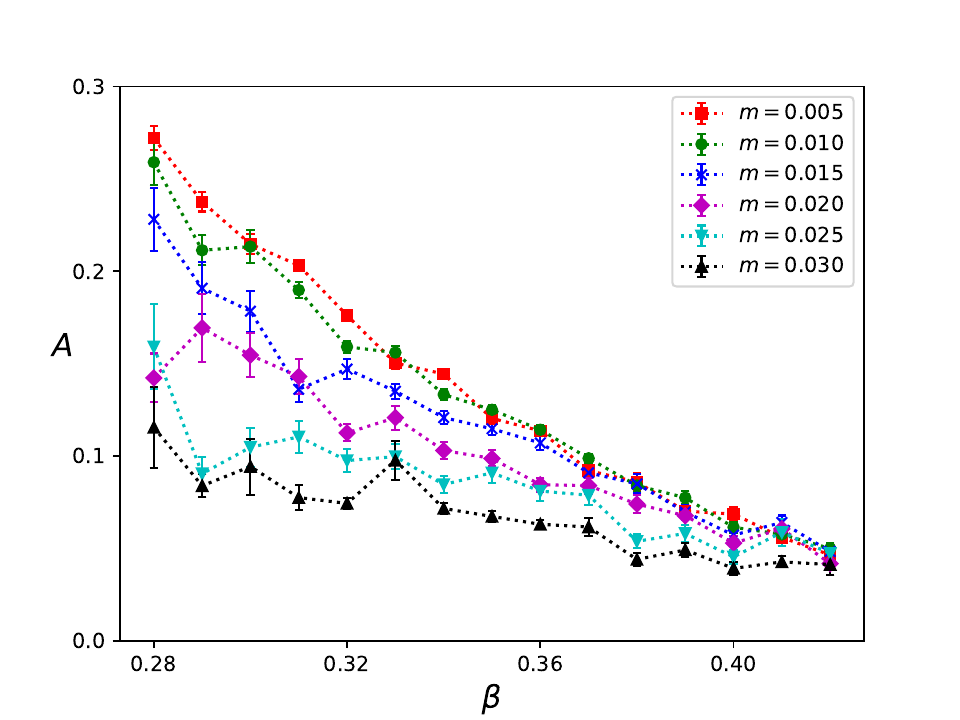}
\includegraphics[scale=0.48]{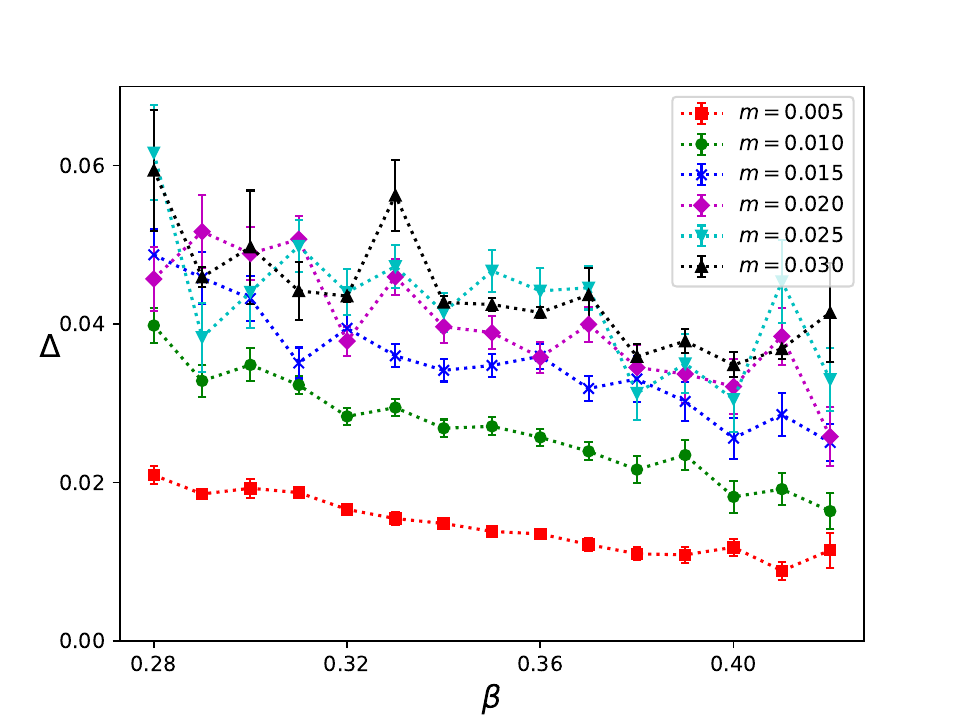}
\end{center}
\caption{Fit parameters $A$ (left panel) and $\Delta$ (right) resulting from
the exponential Ansatz (\ref{eq:extrapolate}) applied to $L_s\in\{24,32,48,72,96,120\}$.}
\label{fig:extrap}
\end{figure}
The right panel of Fig.~\ref{fig:delta} compares the fit with data for three
representative couplings at $ma=0.005$, and 
Fig.~\ref{fig:extrap} shows the fit parameters extracted using 
data from 6 values of $L_s\in[24,120]$. The average $\chi^2$ per degree of
freedom over the 90 parameter sets is 2.15, with no value exceeding 6 save for
one outlier at $\beta=0.31,ma=0.02$ (which does not stand out in
Fig.~\ref{fig:extrap}). Below we shall explore the consequences of removing
some data from the fit. Within fluctuations which increase 
as $A$ decreases and/or $\Delta$ increases (so the overall extrapolation 
decreases in magnitude), the plots show a regular behaviour as
functions of both $\beta$ and $m$. It is interesting that the fit quality
improves as the extrapolation becomes numerically more important, ie. at small
mass and strong coupling. A comparable analysis
for the Shamir kernel is shown in Fig.~2 of \cite{Hands:2020itv}.

\begin{table}[ht]
\begin{center}
{\begin{tabular}{@{}cccccc@{}} \toprule
$L_s({\rm sea})$  & $S_{\rm bose}$ & $\bar\Phi_{24}$ & $\bar\Phi_{32}$ & $\bar\Phi_{120}$ &  $\Phi_\infty$   \\
 \hline
24 &  1.0553(12)  & 0.0814(8) & 0.0916(10) & 0.1582(24) & 0.1821(25)    \\ 
32 &  1.0569(12)  & 0.0798(8) & 0.0905(11) & 0.1480(21) & 0.1644(31)    \\ 
 \hline
\end{tabular} 
\caption{Comparison of ensembles generated  with varying $L_s$ at $\beta=0.34$,
$ma=0.005$.}
\label{tab:24vs32}}
\end{center}
\end{table}
It is important to quantify the uncontrolled and non-unitary approximation
made by taking $L_s({\rm sea})\not=L_s({\rm valence})$, assumed in our analysis
to make negligible
difference.
Table~\ref{tab:24vs32} compares the reference ensemble at $\beta=0.34$,
$ma=0.005$ 
generated with $L_s=24$ and a similar-sized one generated with $L_s=32$.
The values for the auxiliary boson action $S_{\rm bose}$ are compatible, but
higher-order correlations may still be important. We follow the same analysis procedure,
measuring the sequence $\bar\Phi_{L_s}$ and then extrapolating to determine
$\Phi_\infty$. For $L_s({\rm valence})=24,32$ the results are compatible, but by $L_s=120$
a small but significant difference is present, resulting in
$\Phi^{32}_\infty<\Phi^{24}_\infty$. With present resources we are unable to
determine whether this is a fluctuation due to insufficient
statistics (our choice of parameters maximised statistical fluctuations, which
may in retrospect not have been optimal), or a genuine feedback effect. The fitted value
$\Phi_{\rm EoS}=0.1771$
emerging from the equation of state analysis presented below lies in between
but closer to the ensemble generated with $L_s=24$. We can at
least conclude that even if partial quenching introduces significant systematic
errors, they are fairly small.

\begin{figure}[htbp]
\begin{center}
\includegraphics[scale=0.80]{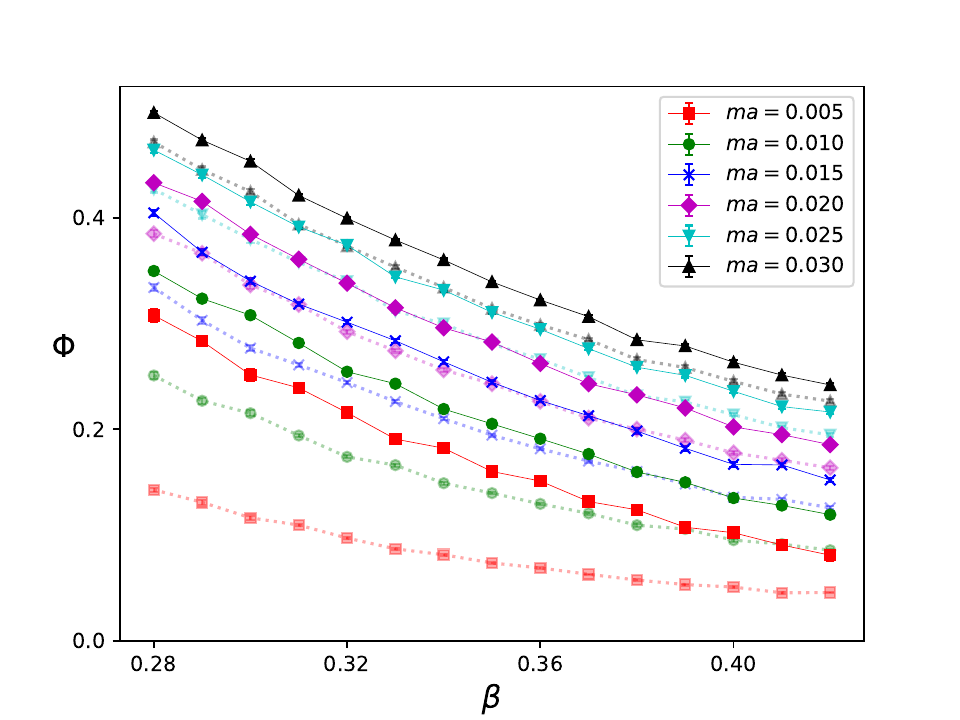}
\end{center}
\caption{Bilinear condensate $\Phi(\beta,m)$ from original unitary simulation on $L_s=24$
(faint symbols, dotted lines), and extrapolated $L_s\to\infty$ (bold symbols,
full lines).}
\label{fig:cond}
\end{figure}
Fig.~\ref{fig:cond} plots the condensate order parameter $\Phi$ in the $L_s\to\infty$
limit together with the original data $\bar\Phi_{24}$ from the RHMC production
run (henceforth we drop the $\infty$ subscript from the extrapolated data).
The extrapolation is considerably more compelling than that obtained with Shamir
kernel fermions (see Fig.~4 of \cite{Hands:2020itv}), for which only RHMC
simulations with $L_s\leq48$ were available, and the partially quenched
strategy was not employed.
For the smallest mass $ma=0.005$ the extrapolation doubles
the size of the signal, underlining the need for a robust 
procedure. Close inspection of the figure reveals that small
fluctuations about the trend in the original data are echoed, but
{\em not\/} magnified in the extrapolated data, suggesting that if needed further accuracy
should be attainable in a future simulation with enhanced statistics. 

\subsection{Equation of State}
On the basis that there is a continuous U(2)-symmetry breaking phase
transiton at some $(\beta_c,m=0)$, we proceed on the assumption that there is a
universal scaling function $\cal{F}$ applicable in the vicinity satisfying
\begin{equation}
m=\Phi^\delta{\cal F}((\beta-\beta_c)\Phi^{-1/\beta_m}),
\end{equation}
where $\delta,\beta_m$ are {\em critical exponents\/} characterising the
universality class of the transition. For the simplest non-trivial case of a
linear scaling function we obtain the following equation of state Ansatz:
\begin{equation}
m=A(\beta-\beta_c)\Phi^{\delta-1/\beta_m}+B\Phi^\delta,
\label{eq:EoS}
\end{equation}
with 5 fitting parameters.
\begin{figure}[htbp]
\begin{center}
\includegraphics[scale=0.48]{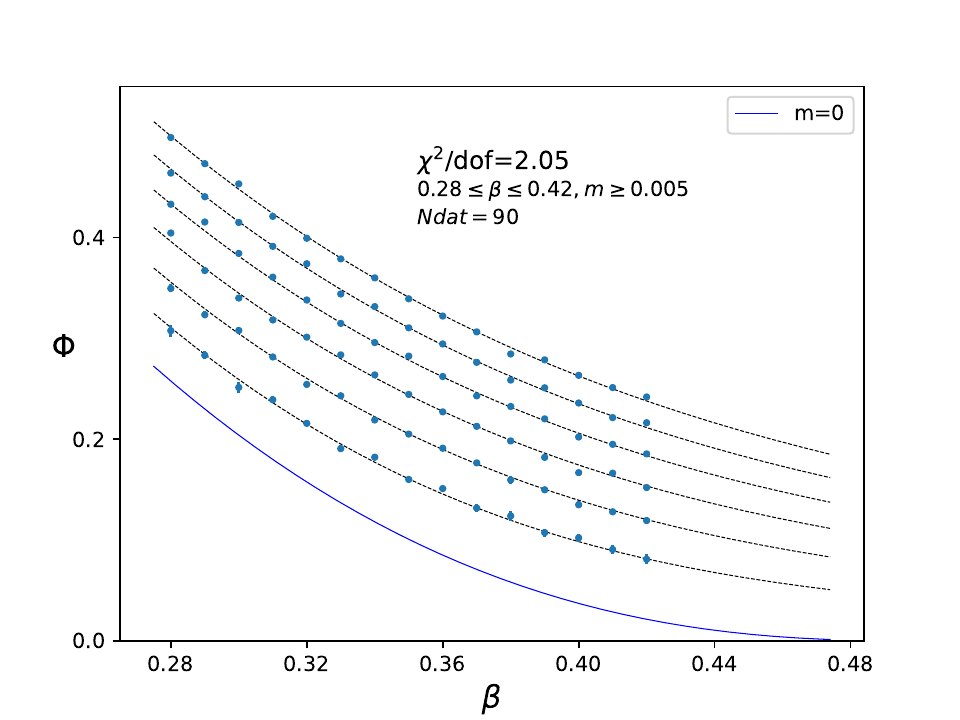}
\includegraphics[scale=0.48]{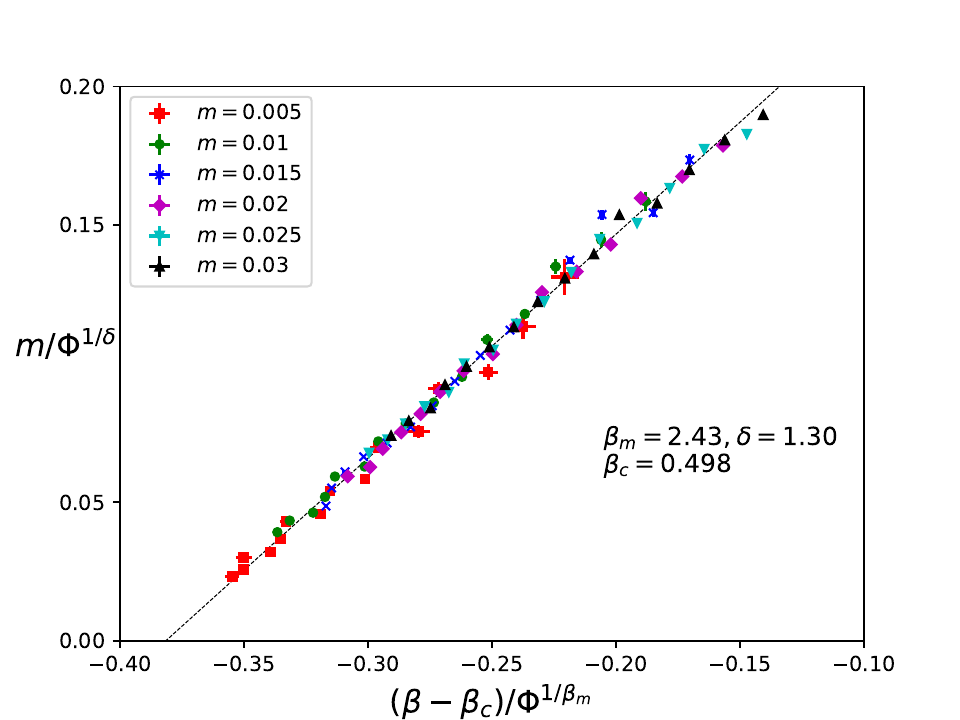}
\end{center}
\caption{(Left) Least-squares fit to the equation of state (\ref{eq:EoS}) using
all 90 data points, with the solid curve showing the $m=0$ limit; 
(Right) Data collapse demonstrating the
near-linear nature of the scaling function ${\cal F}$.}
\label{fig:EoS}
\end{figure}
Fig.~\ref{fig:EoS} shows a least squares fit to the entire dataset
using (\ref{eq:EoS}), yielding $\chi^2/{\rm dof}\simeq2$, which is modest for
a global fit of this nature.
The fit predicts a critical $\beta_c=0.499(15)$, lying well outside the
coupling range explored in the simulation, so that all our datapoints lie in the
symmetry-broken phase. The relatively large fitted value for the exponent
$\beta_m=2.43(15)\gg1$ results in convex constant-$m$ curves approaching the $\Phi=0$ axis with
positive curvature. The right panel of Fig.~\ref{fig:EoS} replots the data to
demonstrate the validity of the linear approximation for ${\cal F}$. The
negative values along the horizontal axis underline that all data lie in the
broken phase. While there is some scatter around the trend line, there is no
sign of any systematic departures. Our quoted values for the critical parameters
are thus
\begin{equation}
\beta_c=0.498(15);\;\;\;\beta_m=2.43(15);\;\;\;\delta=1.300(36).
\label{eq:fitparms}
\end{equation}

The fit is remarkably stable;
we have checked that omitting $\bar\Phi_{24}$ data from the extrapolation,
or data with $\beta<0.30$, $\beta>0.40$ or $ma>0.025$ from the fit makes
negligible difference either to the quality of the fit or the fitted values,
which vary well within the errors quoted in (\ref{eq:fitparms}). Omitting
$\bar\Phi_{120}$ from the extrapolation, which increases the size of the errors
for the lightest mass $ma=0.005$, or equivalently omitting $ma=0.005$ data from
the fit altogether does
significantly alter the fit to $\beta_c=0.453(18)$, $\beta_m=2.06(16)$,
$\delta=1.429(49)$ with $\chi^2/{\rm dof}=2.2$. These data could conceivably be
the most susceptible to finite-volume effects not considered in this
study; accordingly we take these values as conservative lower (upper) bounds for
the exponents $\beta_m$ ($\delta$).

Finally, we apply hyperscaling relations to deduce the value of further
exponents, defined  below in Sec.~\ref{sec:SDE}. Hyperscaling for a
critical system in $d$ dimensions 
is derived under the assumption that there is a single physically-relevant scale,
the correlation length $\xi$: 
\begin{equation}
\delta={{d+2-\eta}\over{d-2+\eta}};\;\;\;
\nu={1\over d}(2\beta_m+\gamma)={1\over d}(2\beta_m\delta-\gamma).
\label{eq:hyperscaling}
\end{equation}
The system (\ref{eq:hyperscaling}) together with the scaling relation
$\gamma=\nu(2-\eta)$ and (\ref{eq:fitparms}) yields the
accompanying predictions
\begin{equation}
\eta=1.61(4);\;\;\;\nu=1.86(13);\;\;\;\gamma=0.73(9).
\end{equation}

\subsection{Susceptibility}
The susceptibility associated with the fluctuating order parameter $\Phi$
is defined by
\begin{equation}
\chi_\ell=-{{\partial^2f}\over{\partial m^2}}={{\partial\Phi}\over{\partial m}}
=V\left(\langle\Phi^2\rangle-\langle\Phi\rangle^2\right),
\label{eq:susc}
\end{equation}
where $f$ is the system's free energy divided by spacetime volume $V$. The mean
square quantity $\langle\Phi^2\rangle$ is estimated using the same noise
vectors as the order parameter, retaining only off-diagonal correlations to
avoid contamination from connected fermion-line contributions.
$\chi_\ell$ is an intensive quantity which may diverge near a critical point.  
\begin{figure}[htbp]
\begin{center}
\includegraphics[scale=0.48]{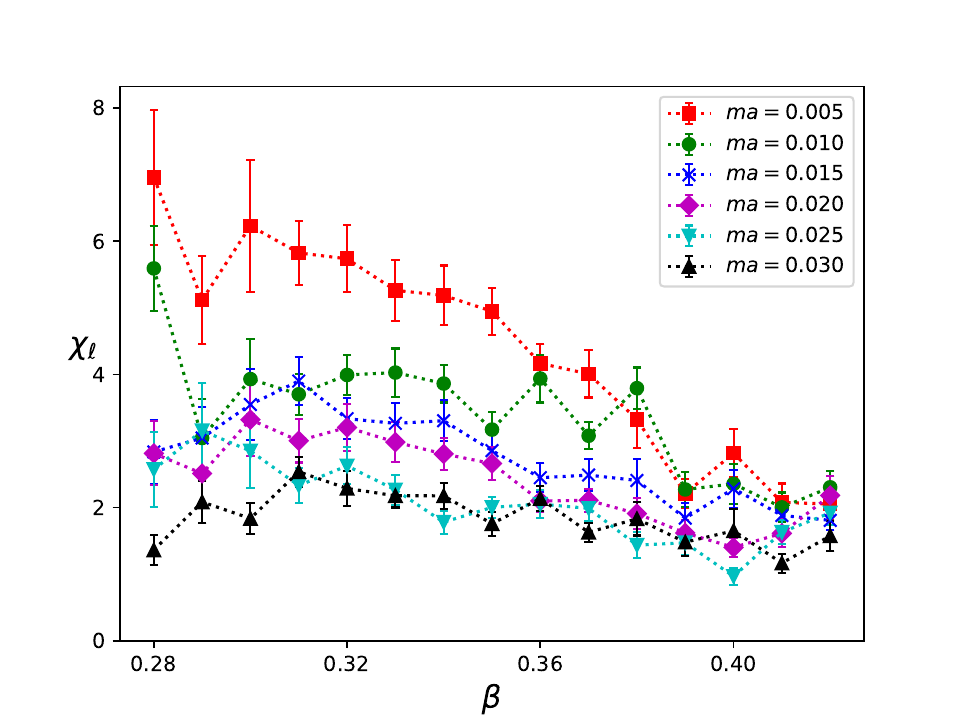}
\includegraphics[scale=0.48]{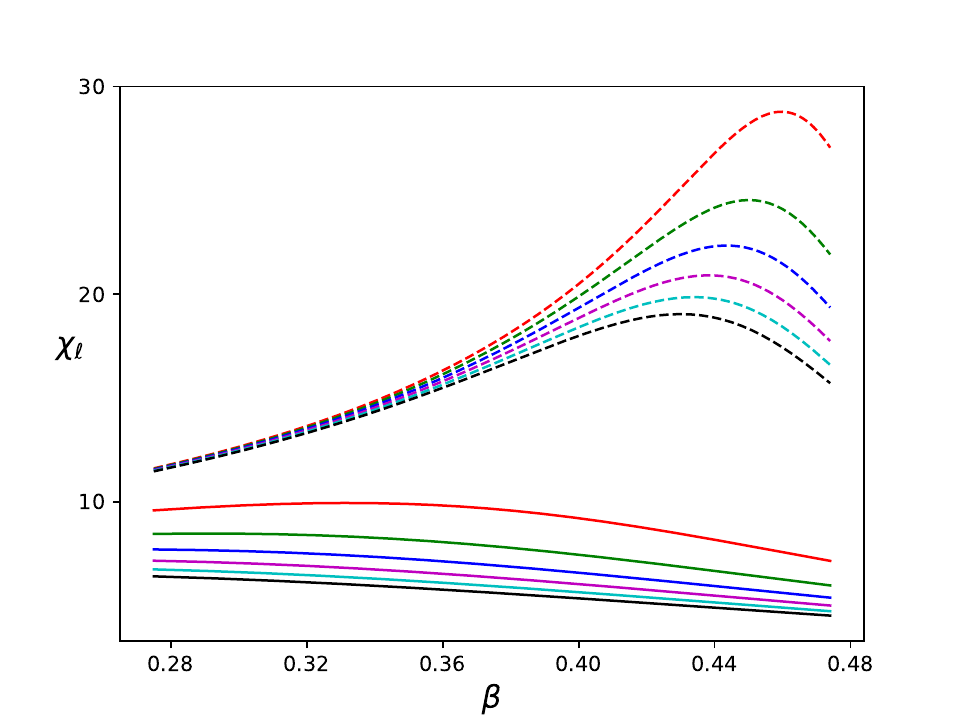}
\end{center}
\caption{(Left) Susceptibility $\chi_\ell$ estimated from the $\bar\Phi_{120}$
dataset; 
(Right) $\chi_\ell$ derived from the fitted equation of state (\ref{eq:EoS})
for the fitted values $m$ (full) and for $m\times10^{-2}$ (dashed).}
\label{fig:susc}
\end{figure}

Experience with simulating the $2+1d$ Gross-Neveu model using domain wall
fermions~\cite{Hands:2016foa}
suggests the $L_s\to\infty$ extrapolation is much more challenging for the
susceptibility than for the order parameter, and the procedure followed in
Sec.~\ref{sec:extrap} yields results which are too noisy to be useful.
Accordingly Fig.~\ref{fig:susc} shows $\chi_\ell$ estimated using measurements
with fixed $L_s=120$, together with an analytic estimate extracted from the
fitted equation of state (\ref{eq:EoS}) using the definition (\ref{eq:susc}).
For these parameters the expected critical peak is rather shallow and
considerably displaced to a coupling stronger than the critical
$\beta_c\simeq0.5$. The right panel of Fig.~\ref{fig:susc} suggests that bare masses
$ma\sim O(10^{-4})$ would be needed in order to see a susceptibility peak develop as
$\beta\to\beta_{c-}$.
The qualitative features of the analytic prediction are to some extent
reproduced by the simulation data, which displays large fluctuations between independent
datapoints consistent with the size of the statistical errors, 
and the order of magnitude of the signal is
comparable, but a quantitative comparison is hardly possible. It should be noted
though, that previous attempts to calculate $\chi_\ell$ using Shamir kernel
yielded unphysical results having an inverted mass hierarchy (see Fig.~9 of
\cite{Hands:2020itv}). Fig.~\ref{eq:susc} represents a considerable improvement,
strengthening our confidence that the partially quenched strategy yields
physically credible
results.

\section{Schwinger-Dyson Interlude}
\label{sec:SDE}
Here we follow the approach to the non-perturbative modelling of the U(2$N$)
symmetry breaking introduced by Itoh {\it et al\/}~\cite{Itoh:1994cr} who
derived results in the strong coupling limit $g^2\to\infty$, and subsequently
refined by
Sugiura~\cite{Sugiura:1996xk} who generalised to finite $g^2$. They
consider an $N$-flavor Thirring model coupled to a St\"uckelberg scalar field
$\phi$, which permits the identification of a ``hidden local symmetry''
$\phi(x)\mapsto\phi(x)+\alpha(x)$,
with the auxiliary $A_\mu$ effectively acting as a U(1) gauge field. The gauge choice
$\phi\equiv0$ recovers the original Thirring model. Next, the full fermion 
inverse propagator including all quantum corrections is parametrised as
\begin{equation}
S^{-1}(p) = iR(p^2)p{\!\!\!/\,}+\Sigma(p^2),
\end{equation}
enabling a set of self-consistent Schwinger-Dyson equations (SDE) for the dressing
functions $R,\Sigma$ to be written. The system only closes if certain
truncations are made; \cite{Itoh:1994cr} chooses to replace the full auxiliary
propagator $D_{\mu\nu}$ and fermion-auxiliary vertex $\Gamma_\mu$ with their forms valid in the large-$N$
limit, namely
\begin{equation}
\Gamma_\mu=-i{g\over\surd
N}\gamma_\mu;\;\;\;D_{\mu\nu}(p^2)\sim{1\over{1-\Pi(p^2)}}={1\over{1+{g^2\over
A_d}p^{d-2}}},
\end{equation}
where $\Pi(p)$ is a scalar function related to the vacuum polarisation tensor 
and $p^{d-2}\equiv (p^2)^{(d-2)/2}$
(since
the underlying $1/N$ expansion is renormalisable in a continuum $d\in(2,4)$ of
spacetime dimensions, it is
helpful to leave $d$ as a parameter). 
Finally, the hidden local symmetry is exploited to define a non-local
gauge fixing functional in which $A_\mu$ and $\phi$ decouple and $R(p^2)\equiv1$, in which case the system
reduces to a single integral equation for the self-energy $\Sigma$:
\begin{equation}
\Sigma(p)=m+{N_c(d)\over 4N}\int_{\mu^{d-2}}^1dq^{d-2}\Sigma(q)
\min\left\{\left(p^{d-2}+{F_d\over g^2}\right)^{-1},\left(q^{d-2}+{F_d\over
g^2}\right)^{-1}\right\},
\label{eq:SDE}
\end{equation}
where we have introduced a bare mass $m$ and an IR cutoff-scale $\mu$
which will eventually be interpreted as an inverse correlation length. All 
dimensionful quantities $\Sigma,p,q,m,\mu,g^2$ in (\ref{eq:SDE}) are defined
in cutoff units, so that the UV limit of the momentum integral is unity. 
For
$d=3$ the
dimensionless constants $A_d,N_c(d),F_d$ have values
\begin{equation}
A_3=8;\;\;\;N_c(3)={128\over3\pi^2};\;\;\;F_3=6.
\end{equation}

The full derivation of (\ref{eq:SDE}), valid for finite $g^2\gg1$, is given in \cite{Sugiura:1996xk};
it requires expanding the gauge-fixing condition consistently to $O(g^{-2})$ and making a
simplified {\em Ansatz} for the polar angular integration which is supported by
a numerical check. Another important simplification is that 
${\rm tr}S_F\propto\Sigma/(q^2+\Sigma^2)$ in the integrand is replaced by the
linearised form $\Sigma/q^2$ valid for small departures from the trivial
solution $\Sigma=0$. It is now possible to recast (\ref{eq:SDE}) as a
second-order differential equation
\begin{equation}
\left[\left(x+{F_d\over
g^2}\right)^2\Sigma^\prime(x)\right]^\prime=-{N_c\over4N}\Sigma(x),
\label{eq:ODE}
\end{equation}
with $x\equiv p^{d-2}$, supplemented by the boundary conditions
\begin{eqnarray}
\Sigma^\prime(p=\mu)&=&0\;\;\mbox{(IR)}\\\left[\left(x+{F_d\over
g^2}\right)\Sigma\right]^\prime\biggr\vert_{x=1}&=&m\;\;\mbox{(UV)}\label{eq:UV}.
\end{eqnarray}

The solution to (\ref{eq:ODE}) is
\begin{equation}
\Sigma(x)={\mu\over\sin\left(\omega\varphi\over2\right)}
\left({{\sigma+f}\over{x+f}}\right)^{1\over2}
\sin\left\{{\omega\over2}\left[\ln{{x+f}\over{\sigma+f}}+\varphi\right]\right\},
\label{eq:solution}
\end{equation}
with $\sigma=\mu^{d-2}$, $f=F_d/g^2$,  and
\begin{equation}
\omega=\sqrt{{N_c\over N}-1};\;\;\;\varphi={2\over\omega}\tan^{-1}\omega.
\end{equation}
For $N=1$ relevant for the current study $\omega\simeq1.823,
\varphi\simeq1.173$.
The UV boundary condition (\ref{eq:UV}) becomes
\begin{equation}
m={{\mu(1+\omega^2)^{1\over2}}\over2\sin\left(\omega\varphi\over2\right)}\left({{\sigma+f}\over{1+f}}\right)^{1\over2}
\sin\left\{{\omega\over2}\left[\ln{{1+f}\over{\sigma+f}}+2\varphi\right]\right\}.
\label{eq:UVBC}
\end{equation}
Finally use the relation for the order parameter
\begin{equation}
{G_d\over{N(1+f)^2}}\langle\bar\psi\psi\rangle=-\Sigma^\prime(x=1)
\end{equation}
with $G_3={16\over3}$ to find
\begin{equation}
{G_d\over
N}\langle\bar\psi\psi\rangle={{\mu(1+\omega^2)^{1\over2}}\over{2\sin\left({\omega\varphi\over2}\right)}}
\sqrt{(\sigma+f)(1+f)}\sin\left\{{\omega\over2}\ln{{1+f}\over{\sigma+f}}\right\}.
\label{eq:psibarpsi}
\end{equation}

To proceed we solve the UV boundary condition (\ref{eq:UVBC}) for $m=0$ by
finding a nodeless zero of the sine function, corresponding to the ground state: 
\begin{equation}
\left({\mu\over\Lambda}\right)^{d-2}=(1+f)e^{2\varphi}\exp\left(-{2\pi\over\sqrt{{N_c\over
N}-1}}\right)-f\sim\xi^{2-d},
\label{eq:xi}
\end{equation}
where we have restored explicit UV cutoff $\Lambda$-dependence to stress the
relation with the correlation length $\xi$. The critical system has
$\mu/\Lambda\to0$ enabling the identification of a critical coupling $g_c^2$:
\begin{equation}
{g_c^2\over\Lambda^{d-2}}=F_d\left(\exp\left[{2\pi\over\omega}-2\varphi\right]-1\right),
\end{equation}
such that the non-trivial solution (\ref{eq:solution}) is the physical one for
$N<N_c, g^2>g_c^2(N)$.
The critical coupling $g_c^2$ diverges as $\omega\to0$, $N\to N_{c-}$. For
$d=3,N=1$ we have $g_c^2/\Lambda\simeq12.04$.

To extract further critical information we use the order parameter
(\ref{eq:psibarpsi})~\cite{Christofi:2007ye}. Away from strong coupling in the critical regime
$\sigma\ll f$ and using dimensional analysis we deduce $\langle\bar\psi\psi\rangle\propto\mu\Lambda^{d-2}$;
however, for $N\sim N_c$ $\sigma\gg f$ and
$\langle\bar\psi\psi\rangle\propto\mu^{d\over2}\Lambda^{{d\over2}-1}$. The
cutoff dependence yields the anomalous dimension of the
composite $\bar\psi\psi$:
\begin{equation}
\gamma_{\bar\psi\psi}={{d\ln\langle\bar\psi\psi\rangle}\over{d\ln\Lambda}}=
\begin{cases} {d\over2}-1 ;&\mbox{strong coupling}\cr
        d-2:&\mbox{otherwise.}
\end{cases}
\end{equation}
In turn this is related to the critical exponent $\eta$ governing critical
correlations of the order parameter fluctuations
$\langle\bar\psi\psi(0)\bar\psi\psi(r)\rangle\propto
1/r^{d-2+\eta}$~\cite{Hands:1992be}:
\begin{equation}
\eta=d-2\gamma_{\bar\psi\psi}=
\begin{cases} 2;&\mbox{strong coupling}\cr
        4-d:&\mbox{otherwise.}
\end{cases}
\label{eq:eta}
\end{equation}
Finally, hyperscaling (\ref{eq:hyperscaling}) enables the extraction of the exponent $\delta$:
\begin{equation}
\delta={{d+2-\eta}\over{d-2+\eta}}=
\begin{cases} 1;&\mbox{strong coupling}\cr
        d-1:&\mbox{otherwise.}
\end{cases}
\label{eq:delta}
\end{equation}
It is important to stress that the limits $N\to N_{c-}$, $g^2\to g_{c+}^2$ do not
commute; in this super-critical regime 
the SDE critical exponents depend on the order of limits and
do not smoothly vary between the two cases set out in
(\ref{eq:eta},\ref{eq:delta}), in contrast to the sub-critical regime of the
gauged NJL model in $d=4$ explored in \cite{Kocic:1990fq}.

Away from the strong coupling limit it is possible to extract exponents by
more direct means. The UV boundary condition (\ref{eq:xi}) can be rearranged
to read
\begin{equation}
\xi=\left({{1+f_c}\over F_d}\right)^{1\over{d-2}}\vert t\vert^{-{1\over{d-2}}}
\end{equation}
with $t=g^{-2}-g_c^{-2}$, whence we identify the correlation length
exponent
\begin{equation}
\nu={1\over{d-2}}.
\end{equation}
Next, rewrite (\ref{eq:UVBC}) with $m>0$:
\begin{equation}
{m\over\mu}=Q\sin(\pi-\varepsilon)\approx Q\varepsilon
\end{equation}
where we assume $\varepsilon\ll1$, so that
$e^{2\varepsilon/\omega}\simeq1+{2\varepsilon\over\omega}$ and
\begin{equation}
\sigma=-{F_d\over{1+f_c}}t+{2\varepsilon\over\omega}{{1+f}\over{1+f_c^{-1}}}\;\;\;
\Rightarrow\;\;\;m=\tilde At\mu+\tilde B\mu^{d-1}=At\Phi+B\Phi^{d-1},
\end{equation}
where the final step applies the scaling $\langle\bar\psi\psi\rangle\propto\mu$
valid away from strong coupling. Comparison with the trial equation of state
(\ref{eq:EoS}) recovers $\delta=d-1$ and the additional prediction
\begin{equation}
\beta_m={1\over{d-2}}.
\end{equation}
Finally the order parameter susceptibility
\begin{equation}
\chi_\ell={{\partial\Phi}\over{\partial m}}\biggr\vert_{m=0}={1\over{A(d-2)\vert
t\vert}}\propto\vert t\vert^{-\gamma}\;\;\;\Rightarrow\;\;\;\gamma=1.
\end{equation}
The exponents derived for $g^2$ finite coincide with those of the $d=3$
Gross-Neveu (GN) model in the large-$N$ limit (see eg. \cite{Hands:1992be}).

\section{Discussion}
\label{sec:disc}
The results reported here show compelling evidence for a U(2)
symmetry-breaking transition in the $3d$ $N=1$ Thirring model at large but
finite interaction strength. The equation of state fits are stable over the
whole parameter range studied, yielding values for two critical exponents
$\delta$ and $\beta_m$, which enables derivation of all others related to the
order parameter assuming the
validity of hyperscaling. The key improvement over earlier studies is much better
control over the $L_s\to\infty$ limit required for U(2) symmetry restoration in
the DWF formulation. Before examining the implications we need to specify some
important caveats. Firstly, our partially-quenched procedure, extrapolating
measurements for varying $L_s$ on auxiliary field configurations obtained with
fixed finite $L_s=24$, is an uncontrolled and non-unitary approximation. As
presented in Sec.~\ref{sec:extrap}, Table~\ref{tab:24vs32}, 
a callibration simulation on $L_s=32$ suggests quenching effects are small, but
with the limited data available at this stage 
a small but statistically-significant effect cannot be ruled out.
Secondly, all results are obtained on a fixed spacetime volume $16^3$; while
previous studies suggest finite volume corrections  are small for similar
models, because the current system looks so different 
we cannot rule out important effects without 
more extensive study. Finally, comparison of the left and right panels of
Fig.~\ref{fig:delta} shows that as $L_s\to\infty$ the residual $\delta_h$ approaches zero more
slowly than the condensate $\bar\Phi(L_s)$ approaches its extrapolated values
$\Phi_\infty$, implying that U(2) symmetry may still not be manifest even for
the extrapolated system we have analysed. This issue also requires further study. 

The system is ``different'' because the fitted values in (\ref{eq:fitparms})
give robust evidence for $\delta<2$, $\beta_m>2$, in marked contrast to all other
studies of the 3$d$ Thirring model using lattice field theory, which yield
$\delta>2$, $\beta_m<1$. Correspondingly, critical correlations appear to be
governed by $\eta>1$, whereas previous studies have yielded $\eta<1$.
For a review see \cite{Hands:2021eyc}, and for a
discussion of $N>1$ with staggered fermions see \cite{Christofi:2007ye}. The equation of state plotted in
Fig.~\ref{fig:EoS} has everywhere convex fit curves, resulting in an unusually large
fitted uncertainty for the critical coupling $\beta_c$. The susceptibility in
Fig.~\ref{fig:susc} shows a shallow peak at $\beta\approx0.3$ only at the lightest mass 
$ma=0.005$, well displaced to the strong-coupling side of $\beta_c$. The fitted
equation of state suggests fermion masses perhaps a hundred times smaller
will be needed before a sizeable peak centred close to $\beta_c$ emerges.
It is safe to state, therefore, that compared to previous lattice work 
the current study reveals a new and
radically different
account of the Thirring model.

One way of articulating the difference is to note that for the first time
lattice field theory simulations reveal a picture qualitatively similar to 
that emerging from self-consistent solution of truncated Schwinger-Dyson
equations, in the sense that $\delta<2,\eta>1,\beta_m>2$.
Sec.~\ref{sec:SDE} reviewed an analytic approach to Thirring criticality
based on SDE solutions. Rather unsatisfactorily, the answers depend on the order
of limits: if 
the strong-coupling limit is taken first, there is an essentially singular
scaling (\ref{eq:xi}) of $\xi$ as $N\to N_{c-}$, with $\delta=1$, $\eta=2$ but other exponents
undefined. If by contrast we study $g^2\to g_c^2$ at fixed $N<N_c$, then 
$\delta=2$, $\eta=\beta=\nu=\gamma=1$. While the exponents predicted in this
case coincide with those of the large-$N$ GN model, they show no
further dependence on $N$, in
marked contrast to GN models where exponents can be calculated
systematically in a large-$N$ expansion~\cite{Hands:1992be}.

How credible is this? Perhaps $N$-independence is a consequence  of
the vanishing of higher-order corrections to the leading-order vacuum
polarisation contribution to the auxiliary propagator, which has been argued
to be essential for the $1/N$ renormalisability of the model~\cite{Hands:1994kb}.
Alternatively, perhaps the $N$-independence is an artefact of the SDE truncations
applied, and a more complete solution might reveal a scenario qualitatively
similar to \cite{Kocic:1990fq}, with exponents varying smoothly 
along the critical line in the $(g^{-2},N)$ plane, ending in an
essential singularity at $(0,N_c)$ where $\delta=1$, $\eta=2$ and $\beta,\nu$
diverge. The fitted values (\ref{eq:fitparms}) seem tantalisingly to favour this
possibility, and strongly motivate a renewed simulation campaign with $N=2$, 
first studied in \cite{Hands:2018vrd}.

\section*{Data Availability}
The simulation data generated for the analysis reported in Sec.~\ref{sec:results} are
freely available at ~\cite{data_access}, along with  links to the production
source code.

\section*{Acknowledgements}
This work used the DiRAC Data Intensive service (CSD3) at the University of
Cambridge managed by the University of Cambridge University Information
Services, 
and the DiRAC Data Intensive service (DIaL 2.5) at the University of Leicester,
managed by the University of Leicester Research Computing Service, each on behalf of
the STFC DiRAC HPC Facility~\cite{dirac}.
The DiRAC
component of CSD3 at Cambridge, and the DiRAC service at Leicester were funded by 
BEIS, UKRI and STFC capital funding
and STFC operations grants.  Additional time on CSD3 was supported by the UKRI
{\em Access to HPC\/} scheme.
DiRAC is part of the UKRI Digital Research Infrastructure.
We are grateful for the help given by Connor Aird and Jamie Quinn 
of UCL's Centre for Advanced Research Computing funded by DiRAC, 
and to Chris Allton and Ryan
Bignell of FASTSUM and Johann Ostmeyer for their support.
The work of JW was supported by an EPSRC studentship.

\end{document}